# Dispersion Limited versus Power Limited Terahertz Transmission Links Using Solid Core Subwavelength Dielectric Fibers

Kathirvel Nallappan, *Graduate Student Member, IEEE*, Yang Cao, Guofu Xu, Hichem Guerboukha, *Graduate Student Member, IEEE,* Chahé Nerguizian, *Member*, *IEEE*, and Maksim Skorobogatiy, *Senior Member, IEEE*

## Abstract

Terahertz (THz) band (0.1 THz-10 THz) is the next frontier for the ultra-high-speed communication systems. Currently, most of communications research in this spectral range is focused on wireless systems, while waveguide/fiber-based links have been less explored. Although free space communications have several advantages such as convenience in mobility for the end user, as well as easier multi-device interconnectivity in simple environments, the fiber-based communications provide superior performance in certain short-range communication applications such as multi-device connectivity in complex geometrical environments (ex. intra-vehicle connectivity), secure communications with low probability of eavesdropping, as well as secure signal delivery to hard-to-reach or highly protected environments. In this work, we present an in-depth experimental and numerical study of the short-range THz communications links that use subwavelength dielectric fibers for information transmission and define main challenges and tradeoffs in the link implementation. Particularly, we use air or foam-cladded polypropylene-core subwavelength dielectric THz fibers of various diameters (0.57-1.75 mm) to study link performance as a function of the link length of up to ~10 m, and data bitrates of up to 6 Gbps at

K. Nallappan is with the Department of Electrical Engineering and Department of Engineering Physics, Polytechnique Montréal, Québec, H3T 1J4 Canada (email: kathirvel.nallappan@polymtl.ca).

Y.Cao, G.Xu, H.Guerboukha, and M.Skorobogatiy are with the Department of Engineering Physics, Polytechnique Montréal, Québec, H3T 1J4 Canada (email: yang.cao@Polymtl.ca, guofu.xu@polymtl.ca, hichem.guerboukha@polymtl.ca & maksim.skorobogatiy@polymtl.ca).

C. Nerguizian is with the Department of Electrical Engineering, Polytechnique Montréal, Québec, H3T 1J4 Canada (email: chahe.nerguizian@polymtl.ca).

Part of the work has been presented in IEEE-CCWC, Las Vegas, U.S.A (2020) [1]

This work was supported by the Canada Research Chair I program and the Canada Foundation for Innovations grant (Project No: 34633) in Ubiquitous THz Photonics of Prof. Maksim Skorobogatiy.

the carrier frequency of 128 GHz (2.34 mm wavelength). We find that depending on the fiber diameter, the quality of the transmitted signal is mostly limited either by the modal propagation loss or by the fiber velocity dispersion (GVD). An error-free transmission over 10 meters is achieved for the bit rate of 4 Gbps using the fiber of smaller 0.57 mm diameter. Furthermore, since the fields of subwavelength fibers are weakly confined and extend deep into the air cladding, we study the modal field extent outside of the fiber core, as well as fiber bending loss. Finally, the power budget of the rod-in-air subwavelength THz fiber-based links is compared to that of free space communication links and we demonstrate that fiber links offer an excellent solution for various short-range applications.

## I. INTRODUCTION

Terahertz (THz) frequency spectrum (0.1 THz-10 THz) holds high promises for many applications that include communications [2], imaging [3] sensing [4] and spectroscopy [4]. In communications, in order to meet the bandwidth demand set by the next generation of wireless systems, a shift in the carrier frequency towards the THz band is unavoidable [5, 6]. THz communications have been already demonstrated in the context free-space wireless links that profit from the presence of several low/modest-loss atmospheric transmission windows [7]. Although there are many advantages of wireless communications including convenience in mobility for the end user, ease in scaling up the network, flexibility of device interconnectivity etc., they also possess many challenges. Particularly, due to high directionality of the THz beams, THz wireless links are known for their high sensitivity to alignment errors, thus requiring careful positioning of the transmitters and receivers [8]. Moreover, reliable communications in non-static environments (ex. between moving objects) require complex beam steering solutions. The situation is further exacerbated in geometrically complex environments (such as inside vehicles, buildings, etc.), where highly complex channel modeling is required. Moreover, free space links have higher chances of eavesdropping thereby increasing the risks for secure communications [9]. Finally, the atmospheric weather conditions such as rain, snow, fog etc. play a major role in affecting the performance and reliability of the wireless THz links. In view of these limitations of wireless THz communications, short-range THz fiber links (~10 m) can offer an alternative solution as THz fibers present a closed highly controlled propagation environment, they can span complex geometrical paths, and they can offer reliable coupling to receiver and transmitter for both static and dynamic applications. One interesting area of applications for THz fiber links is in reliable onboard connectivity and intra-vehicle communications for military and civil transportation. In Fig.1, we show the schematic of various communication modalities within and between the airborne vehicles, as well as place of THz fibers in such applications. For example, a high-speed THz data link with a moving vehicle can be established using a tracking ground station or another vehicle. A high power and high gain transmitter antenna can be used for such a long distance communication [10]. Given high directionality of the THz signals, multiple antenna modules have to be installed on the vehicle surface to cover several possible directions of communication. Next, received THz signals have to be demodulated and interpreted using expensive and environmentally sensitive signal processing units, which should be preferably located deep inside the vehicle.

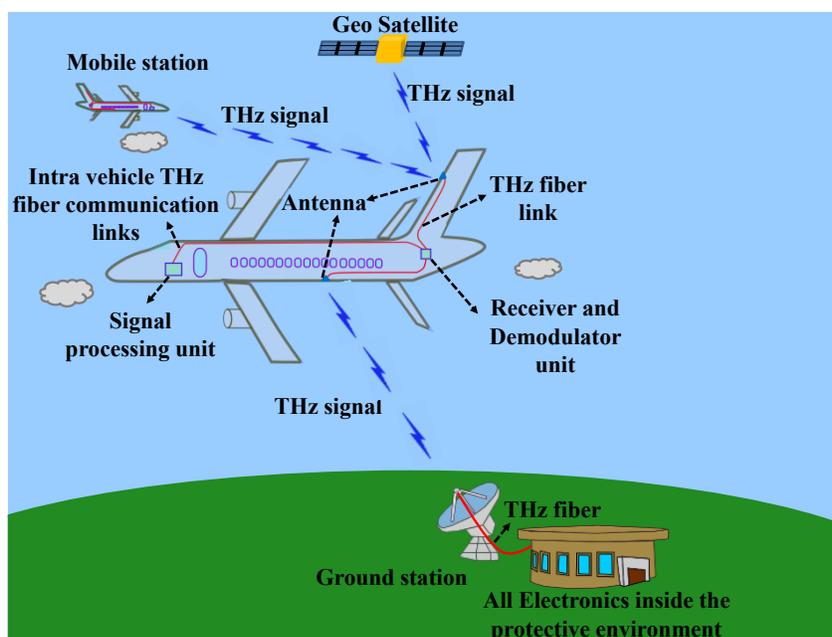

Fig.1. Schematic of the THz wireless and fiber communication links for reliable and versatile intra/inter vehicle communication applications.

This type of scenario where THz signals are detected using multiple antennas, and then relayed over the complex geometrical paths to a central processing unit can profit greatly from the flexible THz fiber links. Another scenario is using THz fiber links for reliable delivery of high-speed data through a partially blocked or geometrically complex areas, which is of importance for hard-to-reach or highly protected environments such as enclosures for aggressive environments (ex. bio-enclosures), protected structures (ex. shelters and bunkers), as well as for intra- and inter-device

THz communications where different parts of the same system can be conveniently linked using flexible fibers. Finally, THz fiber links can be used as a backup solution for short-range wireless communications in case of sudden deterioration of the atmospheric conditions, which can be of particular important for places with harsh weather.

In designing an efficient THz fiber communication link, the fiber parameters such as transmission loss, bend loss, dispersion, coupling strength and ease of handling plays a significant role. Furthermore, the degree of complexity in the fiber fabrication process determine the cost and commercialization opportunities. While the fiber loss and coupling strength limits the communication link distance, the maximum achievable bitrate can also be limited by the fiber dispersion. Therefore, low transmission loss and low dispersion are the primary concerns for the THz fiber designs. We start by reviewing several types of existing THz fibers. The choice of waveguide material is one of the key factors in achieving THz guidance with low loss and low dispersion. In the case of metallic waveguides, the finite conductivity of metallic layers leads to ohmic losses, whereas in dielectric waveguides the loss is mainly due to material absorption. Independently of the materials used, longer THz waveguides (over 1m) are frequently designed to use modes predominantly guided in the low-loss dry air region. Most recently in [11] bare metal wires in air were proposed as open waveguides for 5G communication applications, however such waveguides suffer from high coupling losses and difficulty in mechanical handling due to wire waveguide open structure. Both, low loss, low dispersion and efficient coupling can be achieved using two-wire plasmonic THz waveguides, however longer (over 1m) two-wire waveguides are inconvenient in practice due to challenges in packaging and handling [12]. This is because in the two-wire plasmonic waveguides, the air gap between the two metallic wires should be precisely maintained along the whole fiber length which is difficult to achieve in long fiber links. While encapsulating the two metallic wires within a porous dielectric cladding using fiber drawing offers a solution to the mechanical stability and handling problem, this also leads to addition losses and dispersion due to coupling of a plasmonic mode to the dielectric cladding [13].

Alternatively, by selecting proper dielectric materials with low absorption loss (Teflon, polyethylene, polypropylene, cyclic olefin copolymer to name a few) and engineering the waveguide structure to expel the mode into the low-loss dry air region, highly efficient THz waveguides can be demonstrated [14-32]. In general, dielectric THz waveguides or polymer microwave fibers (PMF) fall under one of the three main categories: hollow core waveguides (anti-resonant reflecting optical waveguides (ARROW) or photonic bandgap (PBG) waveguides) [14-20], porous core waveguides (that use total internal reflection (TIR) or PBG guidance) [21-25] and solid core waveguides (TIR guidance) [26-29]. In the hollow core dielectric tube fibers, the finite thickness of a thin tubular cladding determines the bandwidth of the low-loss ARROW guidance regime, while such waveguides are generally multimode and can support many core and cladding modes. By increasing the size of the hollow core, the propagation loss can be further minimized (in expense to beam quality) as guided modes propagate almost completely in the low-loss air core [14]. Although the cladding modes can be suppressed and the transmission bandwidth can be improved by introducing lossy tubing material such as PMMA, in such fibers one still excites multiple core modes which becomes problematic for long link THz communications due to inter modal dispersion and inter mode interference [16]. Similarly, in the hollow core THz fibers, by arranging alternative layers of high and low refractive index (RI) cladding material (Bragg fibers) or by introducing judiciously designed (periodic or aperiodic) arrays of air inclusions in the cladding (Photonic Band Gap fibers), the loss and the transmission bandwidth can be improved when compared with the tube-based ARROW fibers [4, 15, 17, 20, 33-36]. Moreover, effectively single mode regime can be achieved in long sections of such fibers, which can significantly reduce effective fiber dispersion. Additionally, low loss and low dispersion can be achieved when using spatially variable dense arrays of subwavelength air holes both in the core and cladding regions (porous fibers) [21, 22]. Apart from the circular and hexagonal porous structure [21], honey-comb [37] and rectangular [38] porous geometries were introduced for which the dispersion is comparable to the THz microwires (subwavelength rod-in-air fibers). Finally, porous fibers with graded density of pores have been demonstrated to significantly decrease inter-mode dispersion when operated in the multimode regime [39]. Both hollow core and porous fiber, however, are challenging to fabricate as they rely on precise arrangements of air inclusions in a polymer, glass, or crystalline matrix [36]. Fabrication of most of such fibers involve drawing under pressure a thermo-polymer or glass-based preforms with drilled or 3D printed air-inclusions. Achieving and maintaining target porosity throughout the length of the fiber requires careful calibration and monitoring of the entire drawing process, which is often challenging due to small dimensions of structured preforms used in such drawings. Recently, an alternative method for the fabrication of THz microstructured and PBG fibers was detailed in [36], where over meter-long monocrystalline sapphire fibers were grown directly from the melt using structured dies. Alternatively, such fibers could be 3D printed directly using infinity 3D printing techniques which was recently demonstrated in [40] where authors continuously printed several meters of a wagon wheel highly porous ARROW fiber using Polypropylene.

In the solid core THz fibers, the transmission bandwidth is much larger than that in the hollow core fibers as the propagation mechanism is TIR. However, the transmission loss in such fibers is much higher and is generally comparable to the absorption loss of the fiber material. In order to minimize the transmission loss, one usually resorts to either subwavelength core dielectric fibers which are simple rod-in-air fibers or rod-in-foam fibers [27, 41] or small solid-core photonic crystal fibers (PCF) with porous claddings [42]. The rod-in-air/foam THz fibers with subwavelength size cores offer low loss and low dispersion guidance as large fraction of the modal fields in such waveguides is guided in the low loss air or foam regions [27-29, 41]. In such fibers, scattering from inhomogeneities along the fiber length such as diameter variation, micro and macro bending, material density variation. etc. are the dominant loss mechanisms due to weak confinement in the fiber core [43]. Scattering loss in such fibers can be somewhat mitigated by increasing the fiber diameter and realizing better confinement in the fiber core in the expense of the increased losses due to materials absorption. By choosing low loss plastics for the core materials, as we show in the following, a good compromise can be found, and multi-meter THz fiber links can be realized. Therefore, despite of all these challenges, the rod-in-air/foam subwavelength fibers is a simple, but reliable platform for enabling various short-range THz communication applications. Furthermore, such fibers can be used to fabricate non-trivial signal processing components such as directional coupler, power dividers, band pass filter etc. capable of real-time THz signal processing, which potentially allows building complete transmission/signal processing subsystems using the same base technology [44-46].

As discussed earlier, although, simple rod-in-air subwavelength fibers are easy to fabricate and potentially offer low propagation loss and low dispersion, however, mechanical manipulation of such fibers and their integration into systems is problematic due to significant extent of the modal fields into air [47]. Therefore, for practical applications, such fibers have to be encapsulated in such a way as not to significantly affect the weakly core-bound guided modes, while allowing direct mechanical handling of the fibers. One solution to this problem is a wagon-wheel structure where solid core is suspended in air using several deeply subwavelength bridges [28]. Fabrication of such fibers is, however, challenging due to complexity of the fiber cross-section. Alternatively, solid subwavelength cores can be encapsulated using low-loss, low-RI (~1) dielectric foams, resulting in what we call throughout the paper rod-in-foam fibers (see Fig. 2). In practical terms, high porosity low-loss material foams with deeply subwavelength air cells are virtually indistinguishable from the uniform air cladding, while providing the necessary mechanical stability and ease of manipulation [48, 49]. Although, in principle, the dielectric foams can contribute to additional propagation losses, the effect is negligible for short distances (several meters) and lower THz frequencies (below 200 GHz) for many of the common foams.

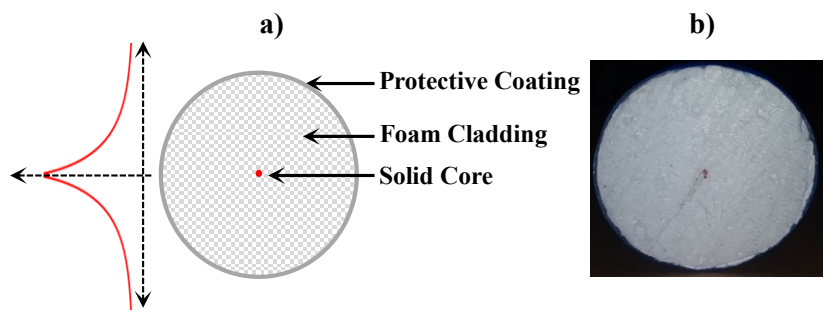

Fig.2. a) Schematic of the rod-in-foam subwavelength THz fiber. Fiber outer diameter is chosen to accommodate ~90% of the power guided by the identical rod-in-foam waveguide with infinite cladding. b) Photograph of the rod-in-foam fiber.

Now, we briefly review some of the recent demonstration of THz communications using THz fibers [50-55]. In [55] an error free transmission of 7.6 Gbps and 1.5 Gbps data over the link distance of 8 m and 15 m has been demonstrated using a hollow core waveguide made of Teflon at the carrier frequency of 120 GHz. The modal loss of the waveguide was 2.5 dB/m. The maximum bit rate at longer link length here was mainly limited by the propagation loss. Similarly, in [56] ortho-mode sub-THz interconnect channel for planar chip-to-chip communications using silicon dielectric waveguide has been investigated. Recently, an 1m expanded porous polytetrafluoroethylene terahertz fiber link have been established together with the resonant tunneling diode integrated with photonic crystal waveguide for 10 Gbps and uncompressed 4K video transmission [57]. For short interconnects, silicon based photonic crystal waveguides were proposed and demonstrated [58-60]. In all these works, however, no in-depth analysis was presented as to the key reasons for the limitations in the transmission length and maximal bitrate in such fiber links. While, the most obvious reason for such limitations is often claimed to be fiber losses, in our following analysis we conclude that fiber dispersion is another major factor that is often overlooked. In fact, we find that depending on the fiber design and operation frequency, one can be in the power-limited or dispersion-limited regime even in the case of short several-meter-long fiber links. From the practical point of view, in the power-limited regime, transmission is possible up to the highest communication bitrate supported by the available hardware even when approaching the maximal link length when signal strength becomes comparable to the noise level. In this regime, the eye diagram opening collapses along a single vertical direction, while no significant degradation in its overall form (skewedness) is observed. In the power-limited regime, negative effects of the modal loss dominate over those due to modal dispersion. In contrast, in the dispersion-dominated regime, even for short fiber links when signal strength is significantly larger than the noise level, one cannot achieve maximal bitrate as allowed by the hardware. In this case, eye diagram shows significant asymmetry and shape contortion due to modal group velocity dispersion and resultant pulse spreading. In the dispersion-limited regime, negative effects of the modal dispersion dominate over those due to modal loss.

In this work we aim at deeper understanding of the limitations of the fiber link quality posed by the combined effects of the modal loss and dispersion. Without the loss of generality, we concentrate on a pure system of rod-in-air dielectric THz subwavelength fiber for a short-range (~10 meters) communication links with up to 6 Gbps data speeds. In fact, the rod-in-air fiber can be used in further studies as a performance benchmark for the more practical fibers such as rod-in-foam and suspended core fibers. In the following, we fix the carrier frequency at 128 GHz, while using fibers of various diameters to realize power-limited or dispersion-limited transmission regimes. The dielectric fibers are made of low loss polypropylene material with three different diameters of 1.75 mm, 0.93 mm and 0.57 mm. Both theoretical and experimental studies are then carried out and comparative analysis of the two is presented. Experiments were conducted using a photonics-based Terahertz (THz) communication system reported in [61, 62].We then demonstrate that the limitation in the error free link distance is mainly due to the modal loss for the 1.75 mm diameter fiber, while for the 0.93 mm and 0.57 mm diameter fibers the link distance is limited due to modal dispersion. By optimizing the decision threshold, an error free ~10m-long link at 4 Gbps is achieved with the 0.57 mm-diameter fiber, while the argument is made for over 10 Gbps fiber links with over 10 m length when designing the fiber to operate near zero dispersion frequency. Furthermore, study of the bending losses of the rod-in-air fibers is presented where we conclude that even relatively tight bends of sub 10 cm-radius can be well tolerated by such fibers. Finally, the power budget of the fiber-based link is compared with that of the free space links and case is made for the strong potential of the rod-in-air fibers in short range communications. To the best of our knowledge, this is the first comprehensive study of all the major limiting factors and comparative advantages that relate to design and operation of the short-range fiber-assisted THz communications links.

## II. THEORY OF ROD-IN-AIR DIELECTRIC THz FIBERS

Many polymers possess almost constant RI and low absorption losses at lower THz frequencies (<300 GHz). Polypropylene (PP), in particular, has one of the lowest losses over the wide THz frequency range (<2 cm$^{-1}$ below 1 THz) [63-65]. Moreover, this material is compatible with 3D printing using cost effective Fused Deposition Modeling (FDM) technique which opens many exciting opportunities in design and manufacturing of various 3D patterned bulk optical components and photonic integrated circuits. Due to importance of PP material for THz application, in our studies we therefore used PP filaments of three different diameters (D=1.75 mm, 0.93 mm and 0.57 mm) as rod-in-air fibers. The filaments having smaller diameters (0.93 mm and 0.57 mm) were extruded using FDM printer. Optical characterization of the fibers were then carried out using an in-house photonics-based THz communication system detailed in [61, 62] that operates at 128 GHz carrier frequency. Complex RI of PP was measured using THz-continuous wave spectroscopy system (see supplementary S:1). Mode analysis of the rod-in-air fibers was carried out using commercial finite element software COMSOL Multiphysics. As the goal of this work is to establish limiting factors in transmission of high bitrate data streams over long distances, therefore modal loss, group velocity dispersion, coupling efficiency and bending losses are the key parameters to model.

### 2.(a) Effective index, Modal Losses, Excitation Efficiency

The normalized electric field distribution |E| of the fundamental HE$_{11}$ mode (normalized to 1W of carrying power) for the PP fibers of different diameters at 128 GHz are shown in fig.3 (a), (b) and (c). The normalized power fraction of the fundamental mode containing within the aperture of a variable diameter centered around the fiber is presented in fig.3(d). Clearly, for the fibers of

larger diameter, the modal field is mostly confined within the fiber, while for smaller diameters the modal fields are strongly present in the low-loss air cladding, which is also the reason for lower absorption losses of smaller diameter fibers. In fig.3(e) we present the effective refractive indices of the fundamental fiber guided modes as a function of the fiber core diameter at 128 GHz operation frequency, the corresponding modal absorption losses (in straight fibers) are presented in fig.3(f). At the carrier frequency of 128 GHz, the fiber operates in a single mode regime up to the fiber diameter of 1.63 mm. For a 1.75 mm fiber one expects excitation of three modes ($HE_{11}$, $TE_{01}$ and $TM_{01}$), while in practice excitation of $TE_{01}$ and $TM_{01}$ does not happen as such modes are incompatible by symmetry with the mode of a WR-6 waveguide (THz emitter waveguide flange) that is centered with respect to the fiber input facet. Thus, all the waveguides used in this work are effectively single mode.

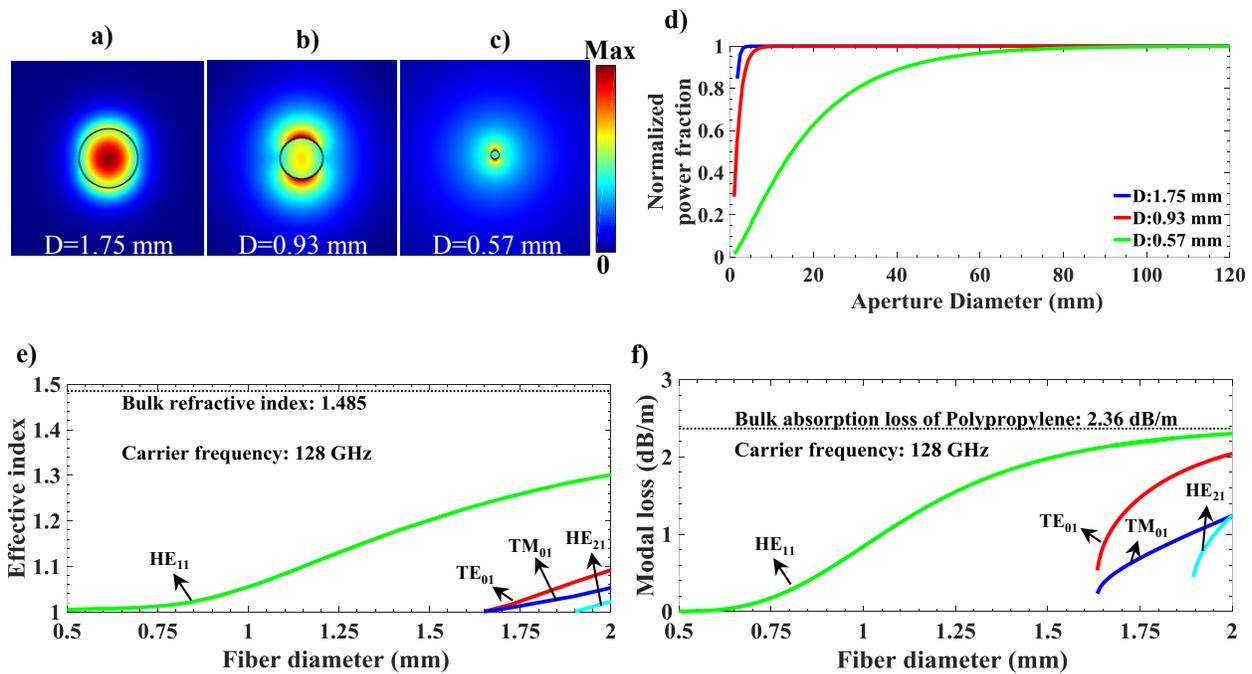

Fig.3 The normalized electric field profile |E| of the fundamental mode at the carrier frequency of 128 GHz. a) 1.75 mm fiber, b) 0.93 mm fiber, and c) 0.57 mm fiber. d) The power fraction of the fundamental mode within the aperture of a variable diameter. e) The effective refractive indices of the guided modes, and f) the corresponding modal absorption losses for the rod-in-air fibers of different diameters at the carrier frequency of 128 GHz. As a reference: the bulk refractive index and absorption loss of the fiber Polypropylene core is 1.485 and 2.3 dB/m respectively at 128 GHz.

We next study excitation efficiency of the fiber fundamental modes using external THz sources. Generally, the excitation efficiency is maximized when the size of the source field distribution is comparable to that of the fiber mode. In our experiments, the subwavelength fibers are butt coupled to the conical horn antenna that is connected to the WR-6 waveguide flange of the THz emitter. The horn antenna is a mode converter with the tapered structure that converts the fundamental TE$_{10}$ mode of a rectangular waveguide to the Gaussian-like mode at the output. Coupling efficiency can be further optimized by properly positioning of the fiber input end inside of the horn antenna. While many coupling scenarios have been considered in the literature [57, 66-68], they all essentially result in a similar coupling efficiency as achieved by a simple free space coupler that uses a single planoconvex lens. Schematic of such a coupler is shown in the inset of fig.4(a). By optimizing the lens parameters (focal length, diameter) so as to match the size of the focused Gaussian beam with that of the fiber mode (or alternatively by matching the numerical apertures of the lens with that of the fiber), one can optimize excitation efficiency of the fiber guided modes. Next, we use transfer matrix theory and a mode matching technique to estimate excitation efficiency of the fiber fundamental mode with the focused linearly polarized Gaussian beam generated by such a coupler/taper at the input facet of a fiber [69, 70]. In fig.4 (a), the excitation efficiency (by power) of the fundamental mode for rod-in-air fibers of three different diameters is presented as a function of the Gaussian beam diameter (1/e$^2$). The maximum excitation efficiency and the corresponding Gaussian beam diameter for all three fibers at 128 GHz carrier frequency is summarized in table 1. Similarly, by using the optimized Gaussian beam size, the fundamental mode excitation efficiency is calculated as a function of frequency which is presented in fig.4 (b). As seen from fig. 4(b), over 90% excitation efficiency of the fiber fundamental mode with the bandwidth of at least 20 GHz is achieved for 1.75 mm and 0.93 mm fiber whereas for 0.57 mm fiber, the bandwidth of ~10 GHz with the excitation efficiency of 80% is achieved.

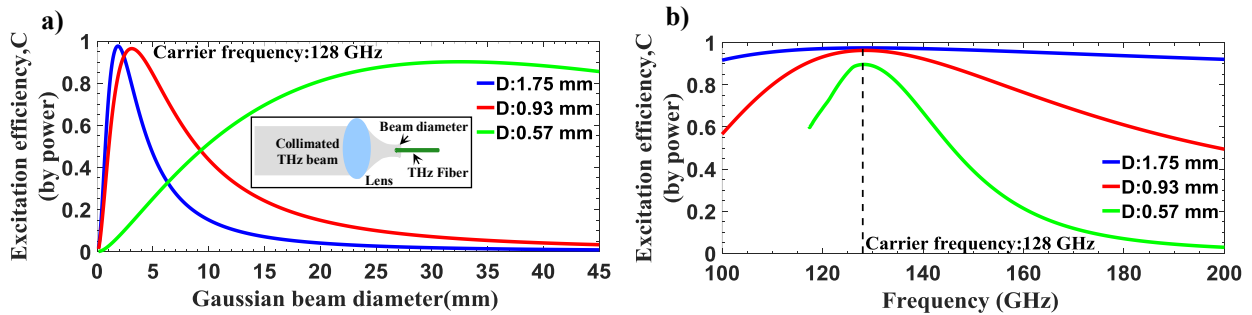

Fig.4. Excitation efficiency by power of the fundamental HE$_{11}$ mode of a rod-in-air fiber of three different diameters. a) Excitation efficiency versus Gaussian beam diameter b) Excitation efficiency as a function of frequency for the optimized Gaussian beam diameter. Inset in fig.4(a): schematic of a simple free space coupler.

| THz fiber | Maximum Excitation Efficiency, C | Gaussian beam diameter |
|---|---|---|
| 1.75 mm fiber | 0.97 | 1.85 mm |
| 0.93 mm fiber | 0.96 | 3.2 mm |
| 0.57 mm fiber | 0.90 | 32.7 mm |

Table 1. The maximum excitation efficiency and its corresponding Gaussian beam size for the fibers of different diameters.

From the data presented earlier in this section we can now estimate the maximal fiber link distance given a ~35dB power budget that is typical for our optics-based THz communication system (for the moment we ignore modal dispersion and bending loss). Thus, received power at the end of the fiber link can be estimated using the following expression:

$$P_r = P_t \cdot C^2 \cdot e^{-\alpha_{wg}L} \qquad (1)$$

where $P_r$ is the power at the receiver end, $P_t$ is the power at the transmitter end, $C$ is the input/output power coupling efficiency per facet (see fig. 4) and $\alpha_{wg}$ is the modal propagation loss by power. The modal loss $\alpha_{wg}$ for all three fibers is obtained from the numerical simulation (see fig.3 (f)) as 2.2 dB/m, 0.62 dB/m and 0.01 dB/m for 1.75 mm fiber, 0.93 mm fiber and 0.57 mm fiber respectively. The transmitter power is -6.6 dBm (~218 μW, which is used in our experiments) and the received power after propagation along the distance L is shown in fig.5.

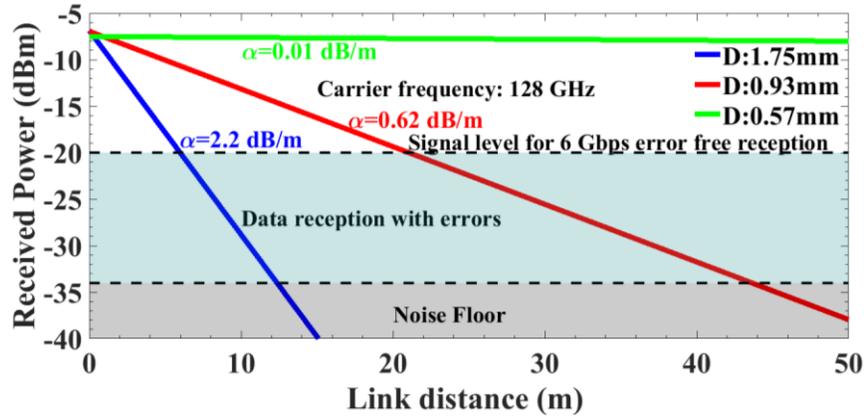

Fig.5. Power budget considerations for the fiber links of variable distance and 6 Gbps data transmission rates used in our experiments. Transmitter THz power is -6.6 dBm (~218 μW). The signal loss level for the error free data transmission is experimentally found at -20 dBm and the absolute noise floor is -34 dBm.

The signal loss level for the error free transmission of data with the bit rate of 6 Gbps is found to be ~-20 dBm (10 μW), while the absolute noise level below which transmission is impossible is found to be at ~-34 dBm (0.4 μW) (see supplementary S:4). From these measurements and from fig. 5 we can estimate maximal distances of the fiber links capable of error-free 6 Gbps data transmission, which are 5 m for the 1.75 mm fiber and 20 m for the 0.97 mm fiber. Note that transmission with errors is possible up to ~10 m in 1.75mm fiber and 43 m in 0.97mm fiber. Finally, we note that while for 0.57 mm fiber the power budget considerations limit transmission distances to ~1 km range, in practice the dispersion and bending loss result in much shorter fiber link distances of several 10's of meters.

*2.(b) Bending Losses*

The bending losses of the guided modes of the rod-in-air THz fibers are calculated using both numerical simulation and analytical approximations. The numerical simulations were carried out using 2D axis-symmetric model in COMSOL Multiphysics. Here, the radiating wave propagates in the azimuthal direction, φ, and the electric field is expressed as

$$E(r, \varphi, z, t) = E(r, z)e^{j(\omega t - \beta r_0 \varphi)} \tag{2}$$

where $r_0$ is the fiber bending radius, while $\beta$ is the leaky mode propagation constant. Computational cell is a rectangle with a circular fiber core positioned at $r_0$ from the axis of rotation, while the other boundaries are perfectly matched layers terminated with a perfect magnetic conductor. Furthermore, we use reflection symmetry with respect to the horizontal plane crossing the fiber center together with the perfect electric conductor or perfect magnetic conductor boundary conditions at the plane to calculate bent fiber modes of two polarizations (electric field at the symmetry plane directed parallel (Y polarization) or perpendicular (X polarization) to the bend axis). Fiber materials are considered lossless for this simulation. Bending losses of the fiber fundamental mode $\alpha_{bend}$ in [dB/m] are computed for the bend radii *R* in the range of 4 mm – 30 mm. The calculated and fitted values of the bending loss are presented on the logarithmic scale in fig.6 (a) as solid and dashed curves for X-and Y-polarizations respectively. In the inset of fig.6, the normalized electric field profile of the bend fundamental mode for the bend radius of 3 cm is presented. For 1.75 mm fiber, the bending loss become smaller than the modal absorption loss in a straight fiber (~2.3 dB/m) for bending radii as small as ~2 cm or larger. At the same time, for the 0.93 mm fiber bending radius should be larger than ~10 cm in order for the bending loss to be smaller than the modal absorption loss (~1 dB/m). We note that optical performance of both fibers (1.75 mm and 0.93 mm) is quite robust with respect to bending as in many practical applications bending radii superior to 2-10 cm can be easily accommodated, however, using fibers of smaller diameters becomes challenging. This is due to the fact that for smaller core diameters, the fundamental mode is only weakly confined to the fiber core leading to higher radiation losses due to bending. For example, as seen from the inset of fig. 6 for the 0.57 mm diameter fiber at 3 cm bending radius, leaky mode of a bent has a much stronger contribution of the radiation continuum to its modal field distribution when compared to the leaky modes of the fibers of larger core diameters. Therefore, while straight 0.57 mm diameter fiber features very low absorption loss <0.1 dB/m, at the same time it is strongly affected by the radiation loss due to micro and macro bending.

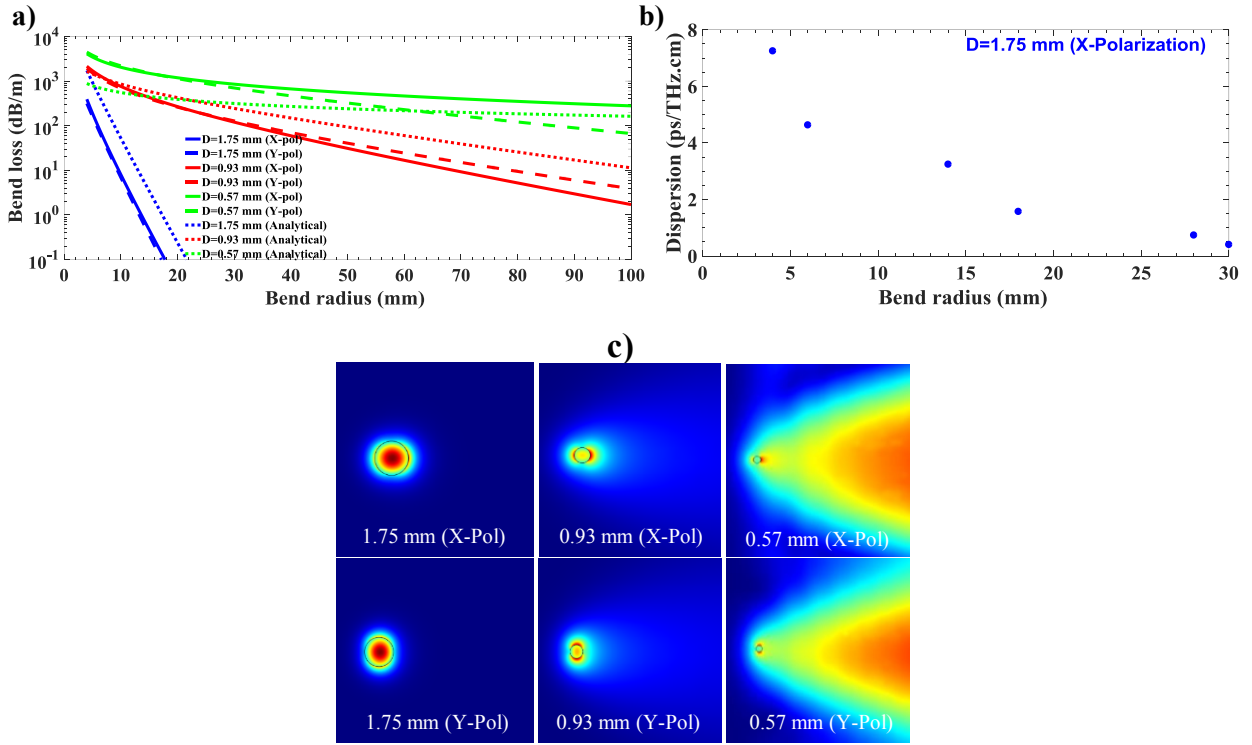

Fig.6. (a) Bending losses of the 1.75 mm, 0.93 mm and 0.57 mm fibers for different bend radius and polarizations. The solid curve corresponds to the X-polarized leaky mode and dashed curve corresponds to the Y-polarization leaky mode of a bend modeled using COMSOL software. The dotted lines correspond to the analytical estimations of the bending loss given by Eq. (3). b). The group velocity dispersion of the fundamental X-polarized leaky mode of the 1.75 mm fiber as a function of the bend radius. c) The field distributions correspond to those of the bend leaky modes for fibers of different diameters and bend radius of 3 cm.

Bending loss can also be estimated analytically using the classical expression for the bending loss of step-index fibers in the regime of weak modal confinement [71, 72]:

$$2\alpha = \frac{\sqrt{\pi}\kappa^2 \exp\left(-\frac{2\gamma^3 R}{3\beta_z^2}\right)}{2\sqrt{R}\gamma^{\frac{3}{2}}V^2 K_{m-1}(\gamma a)K_{m+1}(\gamma a)} \qquad (3)$$

$$\kappa = \sqrt{k_{core}^2 - \beta_z^2}, \qquad \gamma = \sqrt{\beta_z^2 - k_{clad}^2}$$

$$V = ka\sqrt{n_{core}^2 - n_{clad}^2}, k = \frac{2\pi}{\lambda}$$

where, $2\alpha$ is the power loss coefficient, $a$ is the fiber radius, $R$ is the bending radius, $\beta_z$ is the modal propagation constant in a straight fiber and $K$ is the modified Bessel functions where $m$ is the azimuthal mode number corresponding to the subscript $LP_{mn}$. As $HE_{11}$ corresponds to the $LP_{01}$ mode within scalar approximation we use $m = 0$ in Eq. (3). The propagation constants $\beta_z$ of the fundamental modes at 128 GHz for all three fibers are obtained from numerical simulations using COMSOL. Estimated bending losses using analytical expression (3) are presented in fig.6 (a) as dotted lines and show reasonable correspondence with the losses of the bend leaky modes described earlier.

### 2.(c) Modal Group Velocity Dispersion and Maximal Bit Rate Estimation

Next, we study modal group velocity dispersion and maximum error-free bit rate for the three fibers assuming a 10 m-long fiber link. The link length of 10-m is chosen to be long enough to be of practical importance, while making sure that all the fibers have no more than 25-dB loss (by power) over the link distance. In general, the maximum bit rate in the communication link is limited by the pulse dispersion and propagation loss. The dispersion parameter, $\beta_2 \left( \frac{\partial^2 \beta}{\partial \omega^2} \right)$ and $\beta_3$ $\left( \frac{\partial^3 \beta}{\partial \omega^3} \right)$ are the second and third order derivate of the propagation constant $\left( \beta = \frac{2\pi n_{eff}}{\lambda} \right)$ with respect to the angular frequency $\omega$ which is used to characterize the degree of pulse broadening in fibers. Particularly, considering only the second order modal dispersion, the maximum bit rate '$B$' supported by the fiber of length '$L$' can be estimated using expression (4), which is derived by requiring that ~95% of the power of the broadened pulse form still remains within the time slot allocated to logical "1" [73]. In fig.7 (a), the $\beta_2$ of all the three fibers is presented along with their single mode cut-off frequency $f_{sm}$. We see that dispersion ( $\beta_2$ ) of the 0.93 mm fiber (~40 ps/THz.cm) is much higher than those of the 1.75 mm and 0.57 mm fibers at 128 GHz carrier frequency, thus significantly limiting maximal bit rates for the 0.93 mm fiber. Moreover, dispersion of the 1.75 mm fiber is near zero at 128 GHz, thus promising very high bit rates at this carrier frequency. The estimated bit rates for different fibers and different carrier frequencies are presented in fig.7 (b). At 128 GHz and a 10 m-long fiber link, the maximum error-free bit rates of 4.7 Gbps and 1.2 Gbps are predicted for fibers of 0.57 mm and 0.93 mm diameter.

$$B = \frac{1}{4\sqrt{|\beta_2|L}} \qquad (4)$$

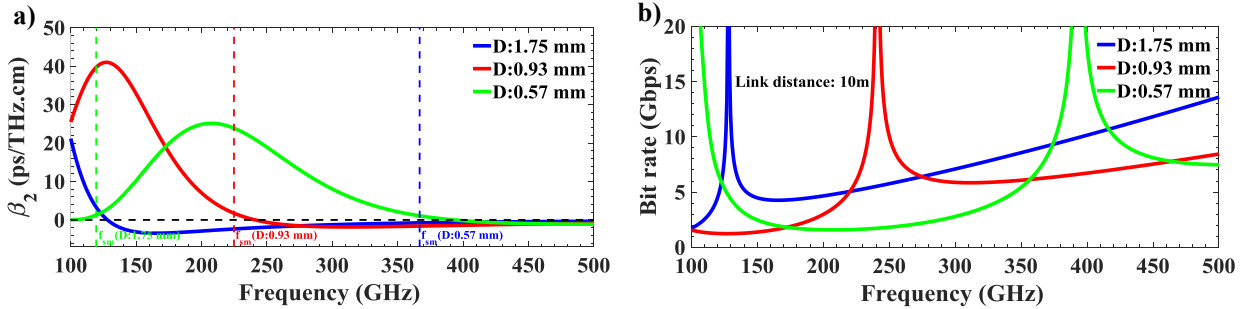

Fig.7. a) The second order dispersion $\beta_2$ of the fundamental mode for 1.75 mm, 0.93 mm and 0.57 mm fibers. The dashed vertical line corresponds to the single mode cut-off frequency of respective fibers. b) The maximum bit rate supported by the fibers in a 10 m link with zero modal loss.

It is important to mention that as 1.75 mm diameter fiber has a Zero-Dispersion Frequency (ZDF, at which $\beta_2 = 0$) in the immediate vicinity of 128GHz carrier frequency used in our experiments. The maximum error-free bit rate $B_{ZDF}$ supported by such a fiber at ZDF is, thus, limited by the third order dispersion $\beta_3$, and can be estimated using expression (5) [73]. Applying Eq. (5) to all the three fibers operating at their respective ZDFs and assuming a link distance of 10 m, we estimate for $B_{ZDF}$ in the 10-20- Gbps range (see Table 2). We thus conclude that, by optimizing the fiber diameter to achieve low dispersion at a given carrier frequency, 10m fiber links can be realized with rod-in-air fibers capable of over ~10 Gbps error-free bit rates per channel, which is sufficient for many practical THz fiber-based communication applications.

$$B_{ZDF} = \frac{0.324}{\sqrt[3]{|\beta_3|L}} \qquad (5)$$

| THz fiber | Zero dispersion frequency (ZDF) | Bit rate at ZDF for a 10 m link |
|---|---|---|
| 1.75 mm fiber | 128 GHz | 9 Gbps |
| 0.93 mm fiber | 241 GHz | 13.8 Gbps |
| 0.57 mm fiber | 393.5 GHz | 19.2 Gbps |

Table 2. The ZDF for 1.75 mm, 0.93 mm and 0.57 mm fibers and their maximal supported bit rates estimated using third order dispersion.

Finally, we note that dispersion of a bent fiber can be different from that of a straight fiber, especially at smaller bending radii. In Fig. 6 b), for example, we show dispersion of the fundamental X-polarized leaky mode of a bent 1.75 mm fiber. At large bending radii dispersion approaches zero (that of a straight fiber), while at bending radii smaller than several cm it grows rapidly and can become as large as ~10 ps/THz·cm. Additionally, ZDW can shift to a somewhat different value in bent fibers, which should be considered when designing fiber links.

## III. EXPERIMENTAL CHARACTERIZATION OF THE ROD-IN-AIR SUBWAVELENGTH FIBERS

### 3.(a) THz Communication System, fiber holding method and principal measurement challenges

The experimental characterization of the fibers was carried out using an in-house photonics-based THz communication system reported earlier in [61, 62].The schematic and the experimental set up of the THz communication system are shown in fig.8 (a) and (b) respectively. Particularly, two-independently tunable distributed feedback lasers (TOPTICA photonics) operating in the infrared C-Band with slightly different center frequencies are used to optically drive the photomixer. The laser beams are combined using a 3dB coupler and intensity modulated (ASK) using an external electro-optic Mach-Zehnder modulator. A baseband signal source of pseudo random bit sequence (PRBS) with varying bit rate from 1 Gbps to 6 Gbps and pattern length of $2^{31}$-1 is used. The modulated laser beams are amplified using an erbium doped fiber amplifier (EDFA) and injected into the waveguide coupled uni-traveling-carrier-photodiode (UTC-PD) photomixer (NTT Electronics) for THz generation. In the receiver section, a zero bias Schottky diode (WR8.0 ZBD-F) is used to detect and demodulate the baseband signal. The baseband signal is then amplified using a high gain low noise amplifier (LNA) and the bit error rate (BER) is measured using the test equipment (Anritsu -MP2100B).

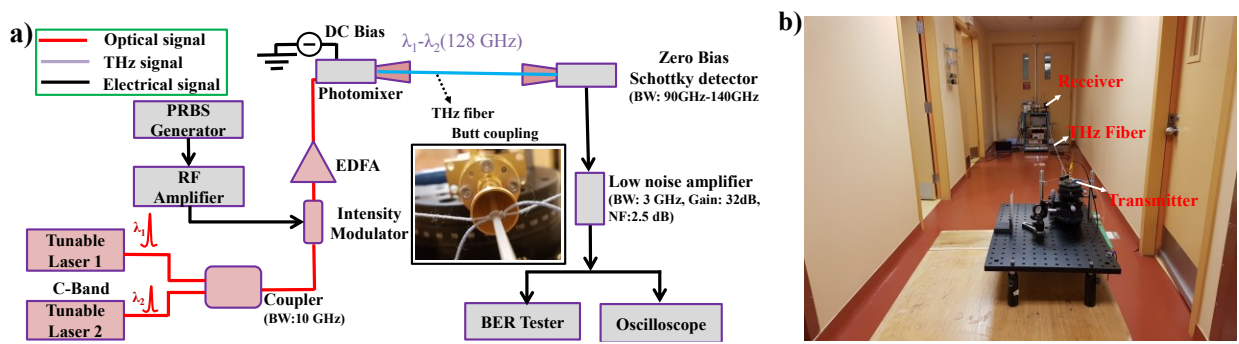

Fig.8. a) Schematic of the photonics-based THz communication system. Inset: Butt coupling of the THz fiber with the horn antenna using fisherman's knot assembly. b) The photograph of the 6 m-long 1.75 mm diameter rod-in-air fiber THz communication link.

The largest rod-in-air fiber used in our experiments is the 1.75 mm fiber which is a commercial 3D printing PP filament (Verbatim) with the diameter of 1.75±0.05 mm. The 0.93mm fiber was fabricated by reducing the diameter of the 1.75 mm PP filament using a 3D printer (Raise3D Pro2). The temperature of the extruder was set to 220˚C and 1 mm nozzle was used for the extrusion. A motorized spinning tool is used to precisely control the diameter of the filament that is extruded from the nozzle by adjusting the drawing speed. Thus, fabricated fiber had 0.93±0.03 mm diameter along the 8-m length. Similarly, a 10 m long 0.57±0.03 mm fiber is fabricated by using the same 1 mm nozzle at increased drawing speed. While characterizing fiber links, a stable and consistent fiber coupling must be used. This is of particular importance for subwavelength fibers that can have significant modal presence in the air cladding. In our experiment, we used butt coupling with the fiber ends judiciously positioned inside the horn antenna. To counter the weight of the free-hanging fiber, the fiber is held tightly using an arrangement of two fisherman's knots made of thin threads and positioned at both emitter and detector ends (see inset of fig.8 (a)). Additionally, rod-in-air fibers do not preserve the in-coupled light polarization state as the light propagates along the fiber. The birefringence is caused by local imperfections in the rod shape (ellipticity, for example), as well as material property fluctuations, which alters the polarization state of the propagating field. Stochastic polarization rotation becomes an issue for fiber links longer than several meters and must be considered while aligning the fiber at the detector side. The optimal alignment is achieved by rotating the fiber end at the detector side until a maximum signal amplitude is recorded.

*3.(b) Measuring fiber propagation loss using cut back technique*

The modal propagation loss of the PP fiber is studied experimentally using the THz communication system and a cutback method, which is then compared to the theoretical values (see fig. 3(f)). Without the loss of generality, here we detail only the measurement for the 1.75mm fiber, while smaller diameter fibers can be characterized in a similar manner. Firstly, the detector electronics is calibrated for direct power estimation from the eye pattern as discussed in the supplementary S:3. Secondly, a 1 m long 1.75 mm fiber is butt coupled at both emitter and detector antenna by direct insertion into the horn antenna linked to the WR-6 hollow rectangular waveguide flange. By fixing the DC bias voltage of the emitter to -2 V, the infrared optical power is increased using EDFA and the eye pattern for 1 Gbps is recorded using the test equipment. When the emitter photocurrent reaches to 4 mA, the eye pattern starts clipping indicating that maximum threshold of the oscilloscope is reached. Therefore, in the following modal loss measurement, the emitter photocurrent is fixed to 4 mA which corresponds to the THz power of -11.5 dBm (~70 μW). Next, the 1.75 mm fiber of length 5m is butt coupled at both emitter and detector antenna. The eye amplitude for three bitrates (1 Gbps, 3 Gbps and 6 Gbps) were recorded. Then the fiber is cut back from the detector side with a step of 0.5 m and the eye amplitude is recorded until the fiber reaches 1 m as shown in fig.9 (a). At each cutback length, the fiber is carefully inserted into the detector horn antenna, while position of the fiber tip is minimally adjusted to achieve the largest opening in the eye diagram. The recorded eye amplitude is then fitted using the form $a \cdot \exp(-bx)$. In fig.9 (b), the estimated THz power is presented in logarithmic scale. It is observed that the modal loss of 1 Gbps, 3 Gbps and 6 Gbps is 2.10±0.37dB/m, 2.00±0.39 dB/m and 1.96±0.40 dB/m respectively which agrees well with the theoretically estimated absorption loss.

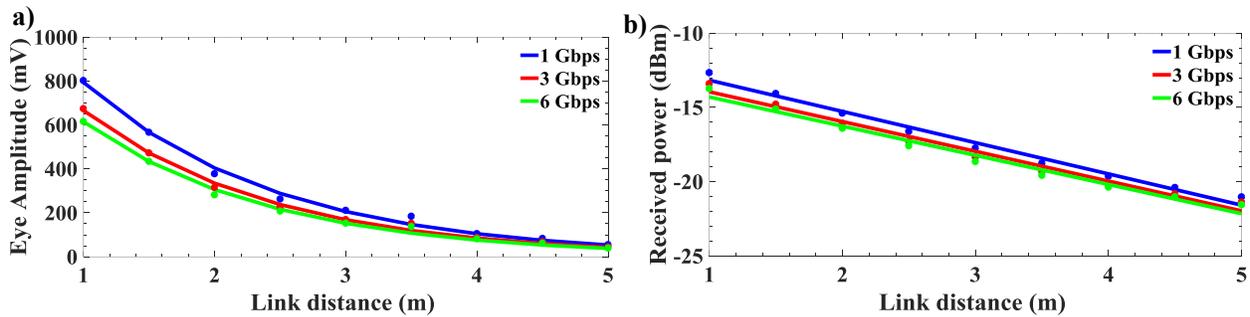

Fig.9. Measuring propagation losses of a 1.75mm fiber using cutback technique. a) Measured eye amplitude for 1 Gbps, 3 Gbps and 6 Gbps signals as a function of the fiber length b) Power loss estimation using detector pre-calibration and recorded eye amplitude.

### 3.(c) Modal field extent in the air

As mentioned earlier, subwavelength fibers can have significant modal presence in the air cladding. In practice, one has to choose the size of fiber cladding (foam, for example) in such a way as to incapsulate most of the modal field. Experimentally, to measure the extent of modal field into air, we place the fiber in the center of a circular metallic aperture of the variable diameter (1 mm − 25mm). The eye amplitude for 1 Gbps bit rate is then recorded as a function of the aperture diameter, from which the fraction of the received THz power is estimated (see fig. 10 (a)) by normalizing with respect to the corresponding value measured for the fully open aperture. From this measurement we conclude that 90% of guided power for the 1.75 mm and 0.93 mm fibers is contained within the circles of ∼ 2 mm and ∼6 mm respectively and thus define characteristic sizes of the claddings to be used with practical fiber designs. At the same time, the corresponding size for the 0.57 mm fiber is theoretically estimated to be ∼45 mm and could not be measured directly due to small size of the used aperture. At the same time, the measured results agree well with the numerical simulations (see fig.3(d)).

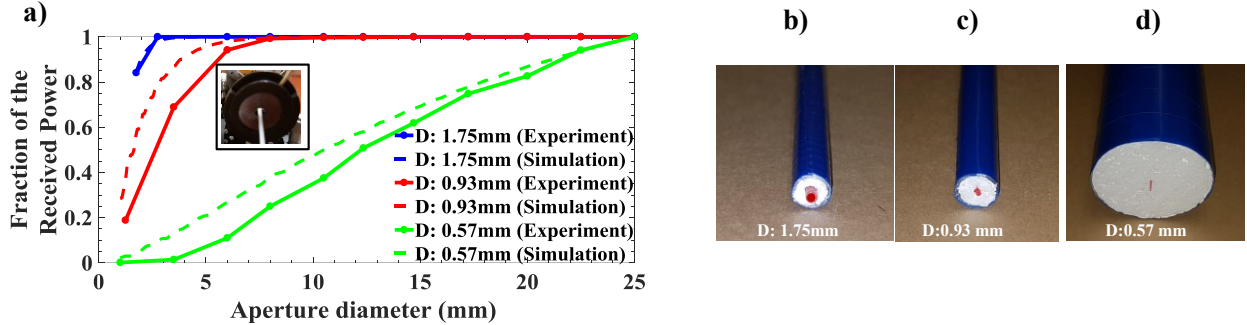

Fig.10. a). Fraction of the modal power inside the aperture of a variable diameter. Inset: Circular aperture centered around the rod-in-air fiber. Photograph of the THz subwavelength fibers with polystyrene foam cladding b) 1.75 mm fiber with 5mm diameter foam cladding (100% of power) c) 0.93 mm fiber with 7 mm diameter foam cladding (99% of power) d) 0.57 mm fiber with 50 mm diameter foam cladding (90% of power).

As mentioned earlier, one of the major challenges posed by the rod-in-air subwavelength fibers is the difficulty in their handling to significant presence of the modal field in the air cladding. To counter this problem, and to enable easy handling and manipulation of such fibers in practical installations, one can insert the subwavelength THz fiber core for example in a circular/square shaped low loss foam that features refractive index close to that of air. In our experiments we realized some of such fibers using polystyrene foams with refractive index of 1.0104 and losses <1 dB/m[12]. In fig.10 (b) (c) (d), we present photographs of several THz subwavelength fibers with fiber diameters of 1.75 mm, 0.93 mm and 0.57 mm surrounded by the polystyrene foam cladding of size 5 mm, 7 mm, and 50 mm respectively, which are chosen from fig. 10 (a) to guarantee that at least 90% of the modal power is confined within the cladding.

## IV.  BIT ERROR RATE MEASUREMENTS

The bit error rate (BER) measurements were carried out to study the fiber link performance under the laboratory environment. In the measurements, the emitter photocurrent is set to 7.5 mA which corresponds to the THz power of -6.6 dBm (~218 μW). A non-return to zero (NRZ) pseudo random bit sequence (PRBS) with the bit rates between 1 Gbps and 6 Gbps and a pattern length of $2^{31}$-1 were used as a baseband signal. For the target BER of $10^{-12}$ (error-free detection), the duration of a single measurement was 1/(target BER · bit rate)~1000s. Furthermore, the decision threshold is optimized so that insertion error (digital 0 is mistaken as digital 1) and omission errors error (digital 1 is mistaken as digital 0) are approximately the same.

*4.(a) BER measurements for the 1.75 mm and 0.93 mm fibers at 8 m link length*

Firstly, we identify the maximal fiber length of the 1.75 mm fiber for BER measurements in our system. For that we consider the eye pattern and observe that beyond 8 meters of fiber length, the eye amplitude becomes comparable to the 0 and 1 noise levels, thus resulting in impractical BER values. Therefore, the maximum link length is fixed at 8 m. Similar to the modal loss measurement, an 8-meter long 1.75 mm fiber is butt coupled to the emitter and detector units using fisherman's knots to hold the fiber in place. First, the BER measurement is carried out for the 1.75 mm fiber by varying the bit rate from 1 Gbps to 6 Gbps. At each bit rate, the decision threshold is optimized so that both insertion and omission errors are the same.

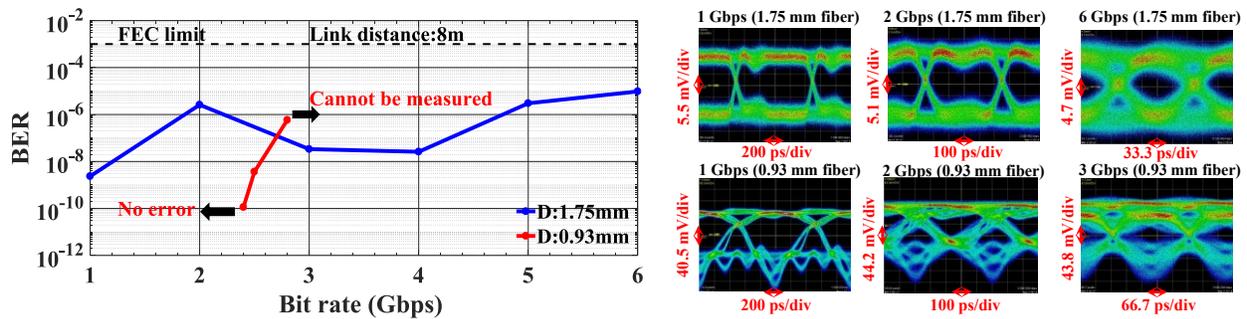

Fig.11.Measured BER versus bit rate for the 1.75 mm and 0.93 mm fibers, and the link length of 8 m. Inset: eye patterns for the two fibers at various bit rates.

The total BER of the 8 m link is presented in fig.11 (blue solid line). Similarly, for the purpose of comparison, the BER measurements of the 8 m-long 0.93 mm fiber link is carried out. In our measurements, we did not observe any error (for optimized decision threshold) for the bit rates below 2.4 Gbps, within the measurement duration. At the same time, we were not able to measure BER for the bit rates beyond 2.8 Gbps due to high group velocity dispersion of the guided mode. The BER recorded for the bit rates between 2.4 Gbps and 2.8 Gbps is presented in fig.11 (red solid line). The inset in fig.11 shows the recorded eye patterns for both the 1.75 mm and 0.93 mm fibers.

From the eye patterns we can judge the effects of modal loss and modal dispersion on the fiber link performance. Thus, in the case of 1.75 mm fiber, the eye amplitude is much smaller than that for the 0.93 mm fiber indicating that 1.75 mm fiber link performance is limited by the modal absorption loss. The error-free operation for the 1.75 mm fiber is observed for the link lengths shorter than 5 meters (Input THz power is ~218 µW) and bit rates of up to 6 Gbps (currently limited by the THz communication system and the input THz power is ~218 µW (-6.6 dBm)). At the same time, even at the link distance of 8 m, the measured BER of $10^{-5}$ is well below the forward error correction (FEC) limit ($10^{-3}$). As the eye patterns for the 1.75 mm fiber stay relatively symmetric even for longer fiber links (8 m) and at high bit rates (6 Gbps), we believe that such a fiber can support at least 9 Gbps up to 10 m (as predicted theoretically), by compensating modal absorption losses with higher input powers (above ~550 μW (-2.6 dBm)). In contrast, for the case of 0.93 mm 8 m-long fiber, although its absorption loss is much lower than that of the 1.75 mm fiber, however, due to much higher group velocity dispersion in such fibers (~40 ps/THz·cm) the maximal bit rate is limited to only 3 Gbps. This is also confirmed by the shapes of the eye patterns for the 0.93 mm fiber that show significant shape degradation at higher bit rates, while also featuring almost 10 times higher powers of the received signals that in the case of the 1.75mm fiber.

*4.(b) BER measurements for the 0.57 mm fiber at 10 m link length*

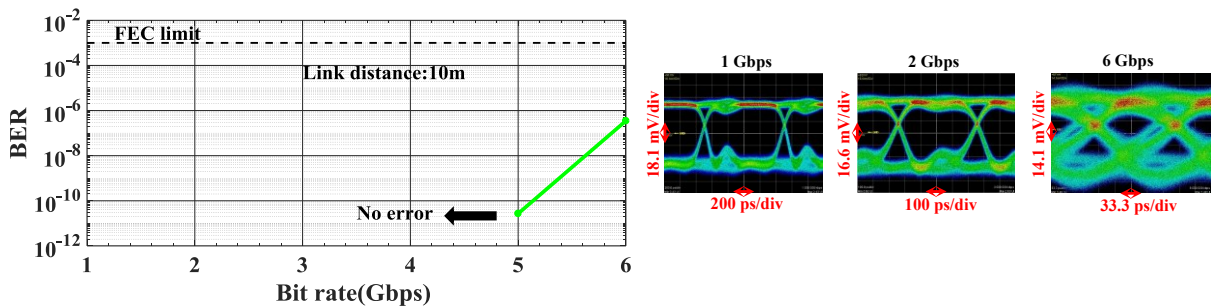

Fig.12. Measured BER versus bit rate for the 0.57 mm fiber and the link length of 10 meters. Inset: eye patterns for 1,2 and 6 Gbps.

We now consider the 0.53 mm fiber that is theoretically predicted to have a very small absorption loss and a relatively small value of dispersion (~4ps/THz·cm). As predicted theoretically, an error free transmission of up to 4.7 Gbps can be achieved using 0.57 mm 10 m-long straight fiber link. Experimentally, we use the same arrangement as discussed earlier, and then conduct BER measurements for data bit rates between 1 Gbps and 6 Gbps and a fiber length of 10 m. Error-free transmission with an optimised decision threshold is observed up to 4 Gbps (measured in steps of 1 Gbps) as shown in fig.13, which is in good agreement with theoretical predictions. The inset in fig.12 presents the eye patterns for 1, 2 and 6 Gbps bit rates. Although propagation loss of the 0.57 mm straight fiber is much lower than those of the 1.75 mm and 0.93 mm fibers, due to considerable modal diameter of the 0.53 mm fiber (~45 mm), a significant portion of the modal power is cut by the horn antenna of 10.8 mm-diameter aperture.

*4.(c) BER measurements for the 1.75 mm fiber and a 90° bent*

Effect of bending on performance of the 8 m-long 1.75 mm fiber is studied using a 90° bend of 6.5 cm bending radius. The BER measurement (see fig. 13) was carried out in the configuration similar to that of a straight fiber detailed earlier, while the fiber bend was realized by suspending the fiber using several knots and holders. For thus chosen bending radius, we observe only a small increase in the measured BER of a bend fiber compared to that of a straight fiber, which is consistent with theoretical observation that bending loss of the 1.75 mm fiber at 6.5 cm bending radius is much smaller than the modal absorption loss, while dispersion of the mode of a bend at this bending radius is also small (<1 ps/THz·cm). We note that performing a reliable measurement of the bending loss using subwavelength rod-in-air fibers is difficult as such fibers cannot be conveniently handled due to significant extent of the modal fields in air. Therefore, we defer a more detailed experimental study of the bending loss to future works which will be conducted using rod-in-foam fibers placed into predesigned bending molds.

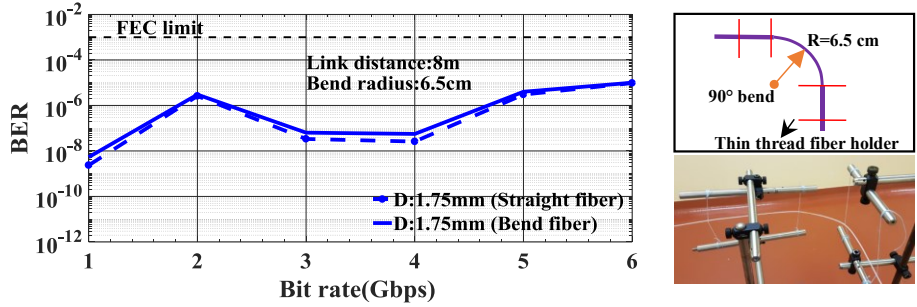

Fig.13. Measured BER for the 90° bent of 1.75 mm fiber with the bend radius of 6.5 cm versus bit rate. The schematic and experimental set up of the bent fiber is shown in the inset.

## V. POWER BUDGET COMPARISON OF THE ROD-IN-AIR THz FIBER LINKS WITH THE FREE SPACE COMMUNICATION LINK

In this section we highlight the advantages of the fiber-based communication links by comparing them with the free space communications assuming a simple ASK modulation scheme. In free space optics, the power received at a distance $L$ from the source is given by:

$$P_r \approx P_t \cdot \frac{D_{RX}^2}{\left(D_{TX} + (L\theta)\right)^2} e^{-\alpha L} \qquad (6)$$

$$\theta \approx 2 \cdot \frac{\lambda}{\pi w_0}$$

where, $P_r$ is the received power, $P_t$ is the transmitter power, $D_{TX}$ and $D_{RX}$ are the aperture size of the source and detector antenna (lens diameter or horn antenna aperture, for example). The angle $\theta$ is a full divergence angle of the beam, $w_0$ is the beam waist size at the transmitter end and $\alpha$ is the air attenuation coefficient. The equation is valid in the far field region i.e. at distances $L > 2 \cdot D_{TX}^2 / \lambda$. A typical power attenuation coefficient $\alpha$ in air at the carrier frequency of 128 GHz is ~6.5 dB/km [74]. The received THz power using fiber link can be estimated using equation (1). As a value for $D_{TX}$ (assuming that $D_{TX} = D_{RX}$) the aperture sizes of a standard horn antenna (10.8 mm), and a lens (2" optics: 50.8 mm) where considered. In free space communication, the received power is mainly limited by the divergence of the propagating beam.

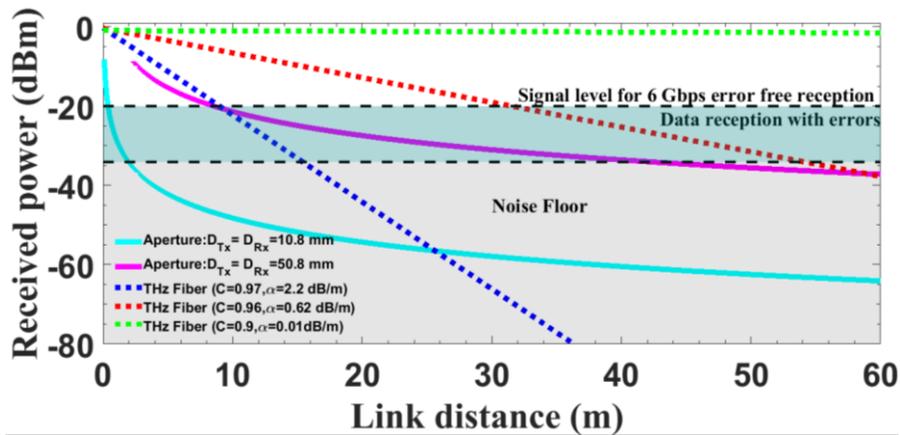

Fig.14. Comparison between free space and rod-in-air fiber-based THz communication links at 128 GHz carrier frequency. The emitter power is set at 0 dBm.

By using large area lens or parabolic reflector antenna, it is possible to collect most of the transmitted energy. However, using large collecting optics is not favorable in many space limited applications. For the emitter power of 0 dBm (1 mW), the received power after distance $L$ for both fiber and free space links are shown in fig.14. From this figure we see that at shorter distances fiber-based links are superior to free space beams in terms of received power due to fast divergence of the THz beams, while at longer distances, free space links are generally superior to fiber links due to smaller atmospheric losses compared to absorption losses of the fiber materials. From the figure we see that when using a relatively small horn antenna (aperture size of 10.8 mm) at both emitter and detector sides, performance of the 1.75 mm fiber-based communication link is superior to that of a free-space link in terms of the received power for the link lengths of up to 24 m. However, at this link distance, the received THz power is well below the noise floor (-34 dBm) for both free space and 1.75 mm fiber-based links. Moreover, even when using relatively bulky 2" optics (lens) for collimating and collecting the THz beam in free space, the performance of 1.75 mm fiber is still superior to a free space link of up to 8 m in length. By reducing the fiber diameter, the modal loss can be further reduced thereby increasing the link distance. Alternatively, the noise floor and the minimal power required for the error-free transmission can be further improved as they depend on many factors including the type of modulation, responsivity of the detector, bandwidth and gain of LNA etc. Although the link distance can be improved by using subwavelength fibers of smaller diameters or by better system design, the maximal achievable bit rate in the fiber links is still much lower compared to the free space THz communication links as fibers feature much higher group velocity dispersion than free space. The maximal bit rate in the free space THz links can be estimated by considering the dispersion of dry air ($2.5 \times 10^{-4}$ ps/THz cm) [71].

| Link distance, L | | 8 m | 15 m | 30 m |
|---|---|---|---|---|
| **Maximal bit rate in Gbps** for different link distances, carrier frequency is 128 GHz | Free space | 560 | 410 | 290 |
| | 1.75 mm fiber | 10.5 | 8.4 | 6.7 |
| | 0.93 mm fiber | 1.3 | 1 | 0.7 |
| | 0.57 mm fiber | 5.2 | 3.8 | 2.7 |
| **Required effective transmitter power in dBm** so that the received signal power is above the minimal -20 dBm level required for the 6 Gbps error-free transmission | Free space ($D_{TX} = D_{RX} = 10.8$ mm, $\alpha$=6.5 dB/km) | 26.3 | 31.8 | 37.9 |
| | Free space ($D_{TX} = D_{RX} = 50.8$ mm, $\alpha$=6.5 dB/km) | -0.57 | 4.9 | 11.0 |
| | 1.75 mm fiber, $\alpha$=2.2 dB/m, C=0.97 | -2.1 | 13.2 | 46.1 |
| | 0.93 mm fiber, $\alpha$=0.62 dB/m, C=0.96 | -14.6 | -10.3 | -0.9 |
| | 0.57 mm fiber, $\alpha$=0.01 dB/m, C=0.90 | -19.0 | -18.9 | -18.8 |

Table.3 The maximal bit rate at different link distances and the required emitter power to result in the -20 dBm signal power (error-free transmission) at the receiver end for both free space and fiber (straight) communication links. The carrier frequency is 128 GHz.

In Table 3 we use both theoretical and experimental data (pertaining to our system) to summarise some estimates for the power budget and maximal bit rates of the fiber-based and free-space THz communication links. Here we use experimentally found -20 dBm as a minimal signal level to achieve the error-free transmission at 6 Gbps bit rates; moreover, since the ZDF for the 1.75 mm fiber is 128 GHz, therefore, third order dispersion $\beta_3$ is used to estimate the maximal bit rate, while for the 0.93 mm and 0.57 fibers we use $\beta_2$ instead.

It is clear from the table 3 that, while the fiber-based communication links offer higher transmitted powers at shorter distances, nevertheless, they also consistently underperform in terms of the maximal achievable data rates compared to the free space links due to higher dispersion. To increase data transmission rates in fiber links, the modal dispersion (as well as modal loss) can be reduced by reducing the fiber diameter to deep subwavelength sizes, however it comes with a higher sensitivity to bending and larger mode diameters. Alternatively, dispersion compensation techniques can be used to compensate for the fiber-link dispersion as we have recently demonstrated using strong hollow-core waveguide Bragg grating [75], however such devices tend to have limited operation bandwidth and further studies are necessary to establish feasibility of this approach. Alternatively, using higher order modulation schemes such as orthogonal frequency division multiplexing (OFDM), the effect of dispersion can be minimized due to smaller bandwidth required at each carrier frequency [76]. This could potentially increase the bit rate several folds in longer THz fiber links.

## VI. CONCLUSION

In this work we presented a comprehensive theoretical and experimental study of simple, yet practical dielectric rod-in-air/foam THz fibers in view of their potential applications in short-range THz communication applications. The THz fibers under study were made of polypropylene and featured three different core diameters of 1.75 mm, 0.93 mm and 0.57 mm. Furthermore, THz communication link performance was characterized with fibers of length 8 m and 10 m as a function of the variable data bit rate 1 Gbps – 6 Gbps at the carrier frequency of 128 GHz. Our main conclusion was that depending on the fiber diameter the communication links were operating either in the power-limited or dispersion-limited regime. Thus, the 1.75 mm fiber featured ~2.2dB/m loss and zero dispersion at the carrier frequency, and it could carry the highest bit rate of 6 Gbps up to the maximal distance of 8 m only limited by the fiber absorption loss, while error-free transmission with such fiber was observed up to 5 m link length. Modal field extent of the core-guided mode into air cladding was only several mm deep due to relatively strong confinement of the modal field in the fiber core. As a result, the 1.75 mm fiber was also well tolerant to bending with virtually no degradation in the link performance when inserting a 90° tight bend of 6.5 cm radius. Further encapsulation of the fiber with polystyrene foam of sub-1cm diameter makes such a fiber an excellent candidate for practical short-range THz communication links due to its ease of handling and installation, as well as good optical properties and tolerance to perturbations such as bending. Similarly, the 0.57 mm straight fiber featured very low absorption loss ~0.01dB/m, and a relatively small dispersion of ~3 ps/THz·cm. The resultant performance was similar to that of a 1.75 mm fiber, however, maximal link length was rather limited by dispersion than by the modal loss. As a result, error-free transmission was realized for a 10m link with up to 4 Gbps data rates, while signal strength was considerably higher than the noise level. One of the major disadvantages of this fiber is high sensitivity to bending and several cm-deep penetration of the modal fields into the air cladding, thus making even the foam-cladded fibers inflexible and somewhat difficult to handle. Finally, the 0.93 mm fiber, while featuring relatively small absorption loss of <1 dB/m, also featured relatively high dispersion of ~40 ps/THz·cm, thus significantly limiting the maximal supported bit rate even for a good signal strength. Aa a result, a maximal bit rate of only 2.4 Gbps was demonstrated for an 8 m fiber link. Finally, we compared the THz fiber communication links with free space links that use relatively small focusing optics (up to 5 cm diameter) and concluded that in this case fiber links are generally more efficient in terms of the power budget for short-range communications up to several 10's of meters. Fiber links are also more reliable and easier to install, maintain and reconfigure than free space links, especially in complex communication environments (on-board communications, for example). At the same time, free space communications outperform fiber-based links in terms of maximal bit rate as air features significantly lower dispersion than fiber.

## ACKNOWLEDGMENT

We thank technician Mr. Jean-Paul Levesque for his assistance.

# Supplementary Material for Dispersion Limited versus Power Limited Terahertz Transmission Links Using Solid Core Subwavelength Dielectric Fibers

Kathirvel Nallappan, *Graduate Student Member, IEEE*, Yang Cao, Guofu Xu, Hichem Guerboukha, *Graduate Student Member, IEEE,* Chahé Nerguizian, *Member*, *IEEE*, and Maksim Skorobogatiy, *Senior Member*, *IEEE*

S1: COMPLEX REFRACTIVE INDEX MEASUREMENT OF THE POLYPROPYLENE FILAMENT

The refractive index of the PP fiber is measured using the continuous wave (CW) THz spectroscopy system [1, 2]. The schematic of the experimental set up is shown in fig.1 which is briefly explained as follows. The setup has two Distributed Feedback (DFB) lasers with slightly different center wavelengths and uniform power (~30 mW each) operating in the telecom region. A 50:50 coupler combines and splits the two wavelengths equally into the emitter and detector arm respectively. Two single mode polarization maintaining fibers wounded with piezo actuators which stretches in the opposite directions were connected to both the arms for the measurement of phase. The symmetric arrangement of the fiber stretchers provides the additional path delay as well as uniform disturbance due to any variation in the external environment. The path lengths between the emitter and detector arms were balanced to have a flat phase.

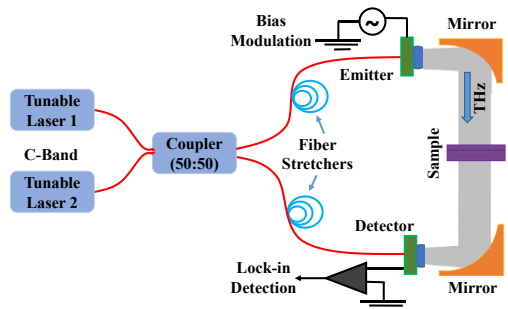

Fig.1. Schematic of the CW THz spectroscopy system for RI measurements.

The generated THz waves which is the frequency difference of the two lasers were modulated using the bias voltage for lock-in detection. The THz beams were collimated and focused into the detector using the parabolic mirrors. The generated photocurrent in the detector were recorded along with the phase information as a function of frequency. The sample is kept in the collimated THz path for the RI measurement. The PP sample for RI measurement is prepared as follows. A circular slab of PP material is fabricated by melting the PP fiber in a crucible at a temperature of 240°C for 35 minutes followed by 20 minutes of cooling. The density of the PP fiber and the fabricated slab after melting is 898.02 kg/m$^3$ and 869.92 kg/m$^3$ respectively with the difference of only <4%. Therefore, it is concluded that the melting process doesn't affect the density of the material and no significant amount of air inclusions were introduced. Three slabs with slightly different thicknesses $d$ were fabricated and then polished on both sides to get a smooth surface. A cutback measurement technique was used to measure the real part of the RI using the CW THz system. Both the amplitude and phase are recorded after removing each slab. The unwrapped phase for different sample (PP) thickness is shown in fig.2(a). Using the phase information, the real part of the RI of PP is extracted using equation (1), where, $n(\omega)$ is the frequency dependent real part of RI, $\phi(\omega, d)$ is the measured phase difference due to presence of a sample and $c$ is the speed of light. The RI of PP is measured as 1.485 in the measured frequency range which is shown in fig.2 (b).

$$n(\omega) = 1 - \frac{\phi(\omega, d) \cdot c}{\omega \cdot d} \qquad (1)$$

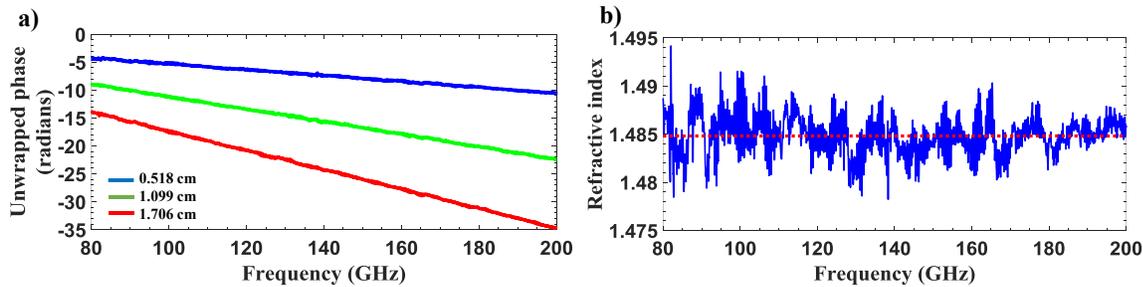

Fig.2.a) Unwrapped phase for different PP slab thicknesses. b) The refractive index of the PP fiber as a function of frequency.

Although the real part of the RI of PP is measured using the CW THz spectroscopy system, extracting the imaginary part accurately requires thick sample (~in meters) due to very low PP material loss. Therefore, we used the modified THz communication system for the loss measurement. The instrumental setups for both the CW THz spectroscopy system and the THz communication are similar (TOPTICA Photonics) except for the additional data modulation unit and detector in the communication system.

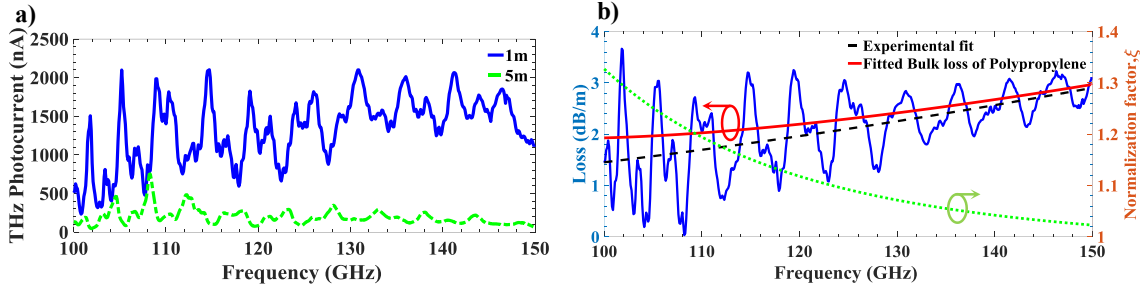

Fig.3.a) THz photocurrent for different fiber lengths. b) Absorption loss of the 1.75 mm PP fiber (blue) as well as inferred bulk absorption loss (black) and a corresponding square fit (red).

In spectroscopy a heterodyning detection scheme is used whereas in communication system, direct detection scheme using Schottky detector is employed. To measure the PP loss, the data modulation unit is disabled, however the bias to the emitter antenna is modulated at lower frequency (12 kHz) for lock-in detection. In the detector section, a trans-impedance amplifier is placed after the LNA where the voltage is amplified and converted to photocurrent which is proportional to the received THz power. In order to measure the loss of the PP fiber, we used the 1.75 mm fiber and the similar fiber holding arrangement is used as for the BER measurement, as well as a metallic aperture at the detector side that was closed around the fiber. By keeping the input coupling undisturbed, the fiber is cut from the detector side during the cutback measurement. The photocurrent of two fiber lengths (5 m and 1 m) were recorded from 100 GHz to 150 GHz as shown in fig.3 (a). The fiber loss is then estimated using equation 2 (see fig.3 (b))

$$Loss \left(\frac{dB}{m}\right) = \frac{1}{(L_5 - L_1)} \cdot 10 \log_{10} \left(\frac{I_5}{I_1}\right) \qquad (2)$$

where, $L_5$ and $L_1$ are the fiber lengths and $I_5$ and $I_1$ are the corresponding THz photocurrents respectively. We note that from (2) one can now extract the bulk absorption loss of the PP material by dividing (2) by a certain frequency dependent normalization function $\xi(f)$. Such function can be computed numerically by assuming a certain frequency independent bulk absorption $\alpha_b$, then finding numerically the corresponding absorption loss of a 1.75 mm rod-in-air fiber $\alpha_f(f)$ made of such a material, and then defining $\xi(f) = \alpha_f(f) / \alpha_b$. Such a function is universal as long as bulk loss used in simulations $\alpha_b$ is small. Thus, extracted bulk absorption loss (black curve in fig. 3 (b)) can then be fitted using second order polynomial (equation 3) where $f$ is the frequency in THz.

$$Loss\left(\frac{dB}{m}\right) = 236 \cdot 31f[THz]^2 - 37.75f[THz] + 3.32 \qquad (3)$$

Finally, bulk absorption loss of Polypropylene at 128 GHz is estimated to be 2.36 dB/m.

## S2: SELECTION OF THE THz CARRIER FREQUENCY

The choice of carrier frequency is determined by the output THz power and responsivity of the detector antenna. In fig.4 (a), the THz power versus frequency measured using a calibrated power meter (PM3-Erickson power meter, Virginia diodes) is presented. Similarly, the corresponding developed DC voltage (without data modulation) is recorded (see fig.4(b)) using the oscilloscope (Anritsu-MP2100B). Both these measurements were carried out independently by keeping the DC bias voltage (-2 V) and photocurrent (7.5 mA) of the emitter antenna constant. A higher DC voltage is recorded for the frequency of 130 GHz. However, we observe a better eye pattern at 128 GHz which is also verified by doing the frequency sweep from 125 GHz to 135 GHz.

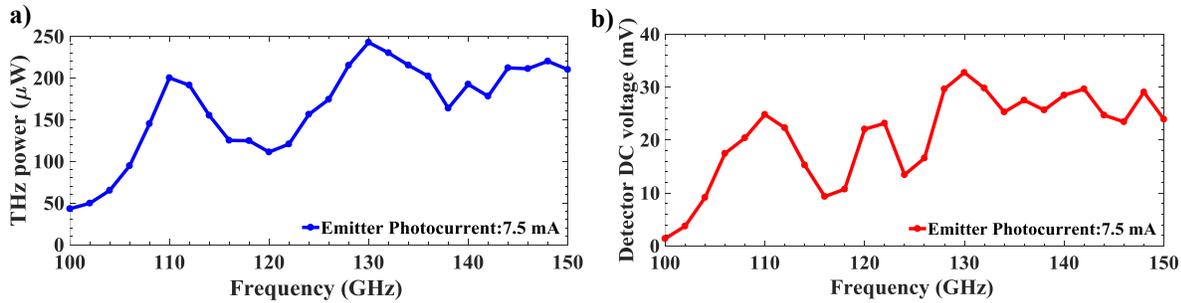

Fig.4. (a) THz output power from the photomixer versus frequency and (b) Developed DC voltage in the detector corresponding to the input THz power shown in (a).

## S3: CALIBRATION OF DETECTOR ELECTRONICS FOR DIRECT THz POWER ESTIMATION

The calibrated calorimetric THz power meter is standard when measuring the absolute power over the broad frequency range. In communications, such THz power meters are integrated with hollow rectangular metallic waveguides for direct coupling with an electronic source. However, for fiber-based links, the absolute power measurement is challenging due to fluctuations caused by coupling consistency issues between the fiber and a waveguide flange. In fact, we find that using horn antenna coupled to a Schottky diode as a detector allows efficient and more consistent optimization of the coupling conditions at the fiber/detector side. Therefore, we first calibrate the detector antenna for direct power measurement using the received digital signal in the communication system. To calibrate the zero bias Schottky diode (ZBD) detector for direct power measurement we proceed as follows. The data modulation unit in the communication system is disabled for the purpose of power measurement. First, the THz output power is measured using calorimetric THz power meter by fixing the output frequency at 128 GHz and the DC bias voltage to -2V. The infrared optical input power to the emitter antenna is varied which is controlled in terms of the photocurrent. The measured THz power as a function of the photocurrent is shown in fig.5 (a).

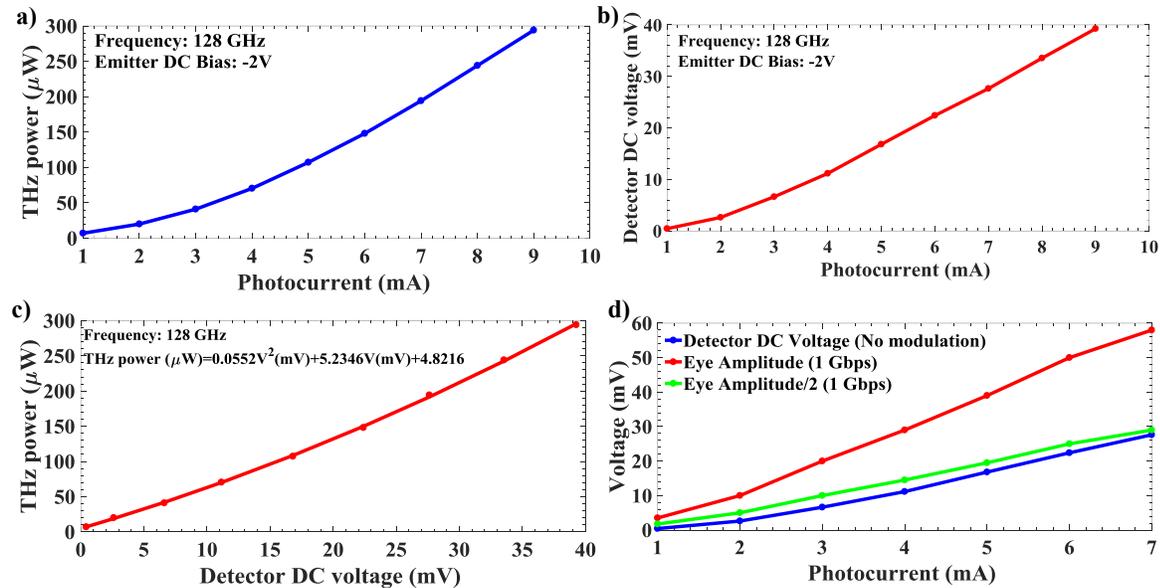

Fig.5 a). Measured THz power and b) Developed DC voltage in the ZBD at the frequency of 128 GHz by varying the input infrared optical power. c). Developed DC voltage in the ZBD versus THz power at the frequency of 128 GHz. d). Relation between the developed DC voltage from the ZBD for the coupled THz signal, eye amplitude and digital one level of the 1 Gbps eye pattern at the carrier frequency of 128 GHz.

Second, the THz emitter is butt coupled to the ZBD by removing the horn antenna attached to it. Similar to the above measurement, the emitter photocurrent is varied which is proportional to the THz output power and the developed DC voltage (without the low noise amplifier) in the ZBD is measured by the oscilloscope with a load resistance of 50Ω. The developed DC voltage as a function of emitter photocurrent is shown in fig.5(b). Then, the photocurrent in fig.5(a) and (b) is replaced with the THz power as shown in fig.5 (c). The measurement in fig.5 (c) is fitted using the second order polynomial:

$$P[\mu W] = 0.0552 \cdot V[mV]^2 + 5.2346 \cdot V[mV] + 4.8216 \qquad (4)$$

where, P is the THz power in microwatts and V is the developed DC voltage in millivolts. Since, the LNA is connected after ZBD during the real-time communication measurements, the additional amplification factor needs to be considered. The LNA is A.C coupled and therefore we cannot measure the amplification of the DC voltage. Therefore, the data modulation unit is now enabled and the 6 Gbps eye pattern is recorded using the high-speed oscilloscope before and after LNA for the THz power of -16.98 dBm (~20 μW). The eye amplitude of the 1 Gbps eye pattern without and with LNA is 9.35 mV and 430.66 mV respectively. The voltage gain in dB is given below.

$$Gain(dB) = 20 \cdot \log_{10}\left(\frac{430.66}{9.35}\right) = 33.26 dB \qquad (5)$$

Similarly, the gain factor of the LNA is calculated as below.

$$Gain\ factor = 10^{\frac{Gain(dB)}{20}} = 10^{\frac{33.26}{20}} = 46.02 \qquad (6)$$

The DC voltage developed in the ZBD by varying the emitter photocurrent (see fig.5 (b)) is recorded when there is no data modulation which is shown in fig.5 (d) as blue solid line. It means that the measured DC voltage is same as logical 1 of the modulated signals (for binary modulation). In our experiments, the digital logical 1 is represented as +1 and logical 0 is represented as -1. Therefore, measuring the eye amplitude and dividing it by 2 is equivalent to the DC voltage measured when there is no modulation as shown in fig.5 (d) (red line and green line). The developed DC voltage in the ZBD is referred here as mean DC voltage recorded when the data modulation unit is disabled. We see a small deviation between the mean DC voltage and eye amplitude/2 which could be due to the amplifier noise and minor discrepancy in estimating the eye amplitude. To summarize the calibration process, the measured eye amplitude is divided by 92.0512 (2*46.0256) and used in the equation 4 to estimate the received THz power.

S4: Noise floor and error-free detection using THz communication system

The following experiment has been carried out to measure the effect of signal level on the BER in our communication system and in order to characterize the noise floor and minimal signal power necessary for error-free detection (BER<$10^{-12}$). In the THz receiver unit, we have used the LNA with the bandwidth of 3 GHz. Therefore, the bit rate in the communication system is limited to the maximum of 6 Gbps (ASK modulation). In this measurement the emitter antenna is butt coupled to the ZBD. The output from the LNA is connected to the test equipment (oscilloscope and BER tester). The DC bias voltage to the emitter antenna is set to -2 V. The emitter photocurrent is varied by increasing the infrared optical power to vary the THz signal power in the system. The eye pattern and BER with optimized decision threshold for the bit rate of 6 Gbps is then recorded. Since the emitter and detector are butt coupled, the received THz power is considered the same as the transmitted THz power which is extracted from fig.5(a).

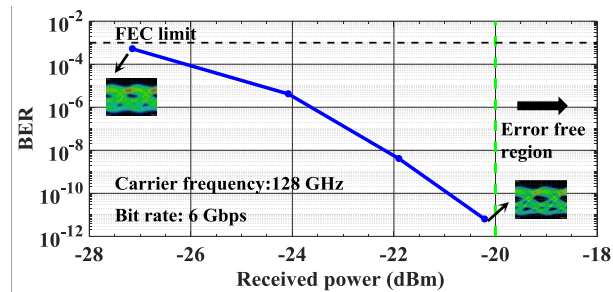

Fig.6. BER measurement in our communication system as a function of the received signal power for the bit rate of 6 Gbps.

From fig.6, for the received THz power of ~-20 dBm (10 μW) and higher, we did not observe any errors within the measurement duration for the optimized decision threshold. Therefore, the signal level of -20 dBm for 6 Gbps is defined as a minimal signal power required for the error-free transmission in our THz communication system.

The noise floor of our THz communication system is measured by decreasing the emitter power, while the mean and standard deviation of one level and zero level of the 6 Gbps data is recorded from the eye pattern. The noise floor is considered as the THz power at which the distance between the mean one and zero levels equals to the average of standard deviations of the one and zero levels. In fig.7, we show the eye pattern of the 6 Gbps data recorded for various emitter powers. The noise floor for 6 Gbps data in our system is ~-34 dBm.

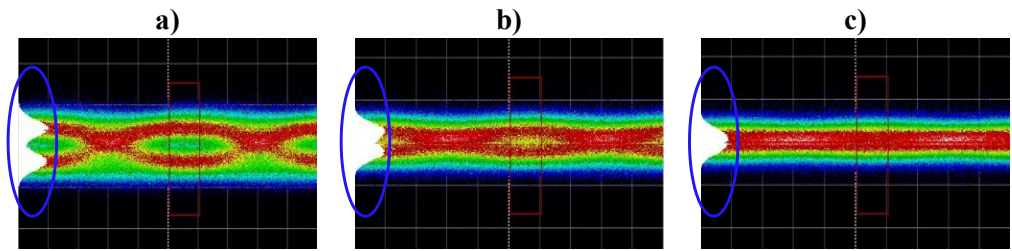

Fig.7. Eye pattern of 6 Gbps data for the emitter power of a). 0.8 μW (-30.96 dBm) b) 0.6 μW (-32.21dBm) and 0.4 μW (-33.97 dBm).